\documentclass[useAMS,usenatbib]{mn2e}

\def \Kepler {\textit{Kepler }}
\usepackage{graphicx}

\title[Characterizing \Kepler Asteroseismic Targets]{Characterizing \Kepler Asteroseismic Targets\thanks{The data used in this paper have been obtained at the Catania Astrophysical Observatory (Italy) the F.L.\ Whipple Observatory, Arizona (USA), and the Oak Ridge Observatory, Massachusetts (USA).}}
\author[J.\,Molenda-\.Zakowicz, D.\,W.\,Latham, G.\,Catanzaro, A.\,Frasca, and S.\,N.\,Quinn]{J. Molenda-\.Zakowicz$^{1}$\thanks{E-mail: molenda@astro.uni.wroc.pl}, D.\,W.\,Latham $^{2}$, G.\,Catanzaro$^{3}$, A.\,Frasca$^{3}$, and S.\,N.\,Quinn$^{2}$\\
$^{1}$ Astronomical Institute of the University of Wroc\l{}aw, ul.\ Kopernika 11, 51-622 Wroc\l{}aw, Poland\\
$^{2}$ Harvard-Smithsonian Center for Astrophysics, 60 Garden Street, Cambridge, MA 02138, USA\\
$^{3}$ INAF - Osservatorio Astrofisico di Catania, Via S. Sofia 78, 95123 Catania, Italy}

\begin{document}

\date{Accepted 2010 November 3. Received 2010 November 3; in original form 2010 July 13}

\pagerange{\pageref{firstpage}--\pageref{lastpage}} \pubyear{2002}

\maketitle

\label{firstpage}

\begin{abstract}

Stellar structure and evolution can be studied in great detail by asteroseismic methods, provided 
data of high precision are available. We determine the effective temperature ($T_{\rm eff}$), surface gravity 
($\log g$), metallicity, and the projected rotational velocity ($v\sin i$) of 44 \Kepler asteroseismic targets 
using our high-resolution $(R > 20,000)$ spectroscopic observations; these parameters will then be used to compute 
asteroseismic models of these stars and to interpret the Kepler light curves. We use the method of cross correlation 
to measure the radial velocity ($RV$) of our targets, while atmospheric parameters are derived using the ROTFIT code 
and spectral synthesis method. We discover three double-lined spectroscopic binaries, HIP\,94924, HIP\,95115, and 
HIP\,97321 -- for the last system, we provide the orbital solution, and we report two suspected single-lined spectroscopic 
binaries, HIP\,94112 and HIP\,96062. For all stars from our sample we derive $RV$, $v\sin i$, $T_{\rm eff}$, $\log g$, 
and metallicity, and for six stars, we perform a detailed abundance analysis. A spectral classification is done for 33 
targets. Finally, we show that the early-type star HIP\,94472 is rotating slowly $(v \sin i = \rm 13\,km\,s^{-1})$ and 
we confirm its classification to the Am spectral type which makes it an interesting and promising target for asteroseismic 
modeling. The comparison of the results reported in this paper with the information in the Kepler Input Catalog (KIC) 
shows an urgent need for verification and refinement of the atmospheric parameters listed in the KIC. That refinement is 
crucial for making a full use of the data delivered by \Kepler and can be achieved only by a detailed ground-based study.
\end{abstract}


\section{Introduction}

The NASA space mission \Kepler\footnote{http://kepler.nasa.gov/} was successfully launched 
in March 2009 with the goal of detecting Earth-size and larger planets by means of the method 
of photometric transits \citep{borucki2009}. 
Having been in orbit since one year, \Kepler has already discovered seven giant planets and delivered 
photometric data for hundreds of stars selected as asteroseismic targets by the 
Kepler Asteroseismic Science Consortium, KASC\footnote{http://astro.phys.au.dk/KASC}
\citep{jcd2007}. 

The aim of the investigation carried out by KASC is a study the internal structure
of the stars by means of asteroseismic methods \citep[see][]{gilliland2010, chaplin2010}. In particular,
the asteroseismic analysis of stars showing solar-like oscillations allows to derive the stars' mean density 
which, when combined with models, yields the stellar mass and radius. The last parameter is one of the key 
characteristics needed to infer the diameters of transiting objects and as such is crucial for the 
study of the planetary systems discovered by \textit{Kepler.}

The asteroseismic analysis, though, is not the only method which can yield precise values of the
radius of the planet-hosting star and thereby the diameter of the planet. As described in detail by 
\citet{sozzetti07}, the stellar density is one of the primary quantities constrained by the analysis of 
the light curve of a transiting body. Therefore, these two independent ways of deriving the stellar
density can be used as a test of the physics involved in the computations. Moreover, as shown by 
\citet{christensen10}, the density derived from the analysis of the light curve can be used to select the 
correct value of the large separation ($\Delta \nu$) of the frequencies of solar-like oscillators for which 
the low signal-to-noise of the data does not allow to detect individual frequencies and makes the 
asteroseismic analysis inconclusive.

Since the asteroseismic modeling requires accurate values of the effective temperature ($T_{\rm eff}$), 
the metallicity and the surface gravity ($\log g$) or an equivalent parameter describing the luminosity
of the star, one of the purposes of the KASC is deriving these values from coordinated ground-based 
spectroscopic and photometric observations as described by \citet{uytterhoeven10} and 
\citet{molenda10}. These efforts are needed because the stated precision of the atmospheric parameters provided 
in the Kepler Input Catalog (KIC)\footnote{http://archive.stsci.edu/kepler/kepler\_fov/search.php} and 
derived from the photometric observations acquired in the Sloan filters \citep{Latham2005}, is 200~K in 
$T_{\rm eff}$, and 0.5~dex in $\log g$ and [Fe/H] which is too low for an asteroseismic modeling. 
Moreover, a significant fraction of \Kepler asteroseismic targets are missing the information on $T_{\rm eff}$, 
$\log g$ and [Fe/H] in the KIC.

This paper is a sequel of a series started in 2006 \citep[see][and the references therein]{catanzaro10}
which aims at deriving atmospheric parameters of stars selected as \Kepler asteroseismic targets by J.M.-\.Z 
and A.F. in the first run of the \Kepler proposals. In this paper, we provide $T_{\rm eff}$, $\log g$,
the metal abundance, as well as the projected rotational velocity $(v\sin i)$ and the radial velocity 
$(V_{\rm r})$ for 44 stars. The information derived by us will be used by the KASC to construct the
evolutionary and asteroseismic models of these stars and to determine their radii as described by
\citet[][]{stello2009}. The information on $v\sin i$ will help select candidates for the long-term 
monitoring of \Kepler asteroseismic targets while the measurements of $RV$ will allow to detect 
new single- and double-lined spectroscopic binaries in the \Kepler field. The latter systems are particularly 
important for the study of the planetary systems since for stars with the orbital solutions the mass of the star
can be computed which allows to determine the mass of the transiting planet.

\section{Observations and data reduction}

\begin{table*}
\begin{center}
\caption{The individual radial velocity measurements of 44 Kepler asteroseismic targets.}
\label{rv-individual}
\begin{tabular}{rrccccrcccc}
\hline\hline\noalign{\smallskip}
 HIP   &KIC-10 &$\alpha _{\rm 2000}$ &$\delta _{\rm 2000}$ & Kep  & HJD      & $RV$         & $\sigma$      &  Instrument &$\rm T_{\rm exp}$ & S/N\\
       &       &                     &                     & mag & +2400000 & [km\,s$^{-1}]$ &[km\,s$^{-1}]$&            & [s]         \\
\hline
 91990 & 8343931 &  18:44:57.58 & +44:21:32.3 & 8.25 &   54656.4941 &    9.75 &  0.83 & FRESCO  &      3600  &   60\\
 91990 & 8343931 &  18:44:57.58 & +44:21:32.3 & 8.25 &   54667.3607 &    9.46 &  0.71 & FRESCO  &      3600  &   95\\
 92962 & 9139163 &  18:56:22.13 & +45:30:25.3 & 8.33 &   54656.5414 &$-$29.16 &  0.17 & FRESCO  &      3600  &   70\\
 92962 & 9139163 &  18:56:22.13 & +45:30:25.3 & 8.33 &   54667.4454 &$-$29.96 &  0.14 & FRESCO  &      2700  &   80\\
 \multicolumn{11}{c}{\dotfill}\\
 97992 &10162436 &  19:54:54.17 & +47:08:43.0 & 8.61 &   54662.5474 &$-$54.30 &  0.17 & FRESCO  &      5400  &   60\\
\hline
\end{tabular}
\end{center}
\end{table*}

The observations were performed at the {\it INAF\,-\,Osservatorio Astrofisico di 
Catania} (OAC), Mt.\ Etna, Italy (63 spectrograms), the Oak Ridge Observatory (ORO), 
Harvard, Massachusetts (162 spectrograms), and the F.L.\ Whipple Observatory (FLWO), Mount Hopkins, Arizona 
(95 spectrograms). We made use also of the archival spectroscopic observations acquired at the 
Multiple Mirror Telescope (MMT), Mount Hopkins, Arizona (10 spectrograms) before it was converted to the 
monolithic 6.5-m mirror. We acquired a total of 288 spectrograms; 6 of our targets were observed more than 20
times, 19 stars, 2-3 times, while for the remaining 19 targets we acquired single spectrograms.

At OAC, we used the 91-cm telescope and FRESCO, a fiber-fed REOSC echelle spectrograph, that allows to 
obtain $R$\,=\,21\,000 spectra in the range of 4300--6800 {\AA} in 19 orders. The gaps between the first 
six orders decrease from 27 to 4 {\AA} going from the red part of the spectrum to the blue. Then, the spectral 
ranges covered by the consecutive orders starts to overlap. The spectra 
were recorded on a thinned, back-illuminated (SITE) CCD with 1024 x 1024 pixels of 24~$\mu$m size, 
typical readout noise of 10 e$^{-}$ and gain of 2.5 e$^-$/ADU. For the reduction of the 
spectrograms, we used the NOAO/IRAF package\footnote{IRAF is distributed by the National 
Optical Astronomy Observatory, which is operated by the Association of Universities for Research 
in Astronomy, Inc.} The reduction included the subtraction of the bias frame, trimming, correcting 
for the flat-field and the scattered light. The spectra were then extracted with the use of the 
{\sf apall} task also provided by IRAF. 

At ORO and FLWO, the spectra were acquired using 1.5-m telescopes and the CfA Digital Speedometers 
which are nearly identical instruments each with the resolving power $R$=35\,000. These echelle spectrographs, 
with photon-counting intensified Reticon detectors, record about 45 {\AA} of spectrum in a single order 
centered at 5187 {\AA}. The spectra were extracted and rectified by means of a special procedure 
developed for these observatories and described by \citet{latham1992}.

\setcounter{table}{1}
\begin{table*}
\begin{minipage}{17.7cm}
\caption{The spectral type and the average radial velocity of the target stars.}
\label{param} 
\begin{center}                
\begin{tabular}{l r c c r l r r r r r}     
\hline                  
\hline            
HIP~     & KIC-10~ & $\alpha_{2000}$                    
         & $\delta_{2000}$                          
         & $Kp$~
         & SpT     
         & $N$
         & Span 
         & $\langle RV \rangle$ 
         & $\chi ^2$ & $\rm P(\chi ^2)$\\
         &         
         & $(^{\rm h}$ $^{\rm m}$ $^{\rm s})$ 
         & $(^\circ$ $^\prime$ $^{\prime\prime})$
         & (mag)& 
         &   
         & (days)
         & (km\,s$^{-1}$)
         &
         &\\
\hline\noalign{\smallskip}
 91990    &  8343931&18:44:57.58&+44:21:32.3&8.25&F5\,V   &   2 &   11 &    $9.08\pm0.54$&  ... & ... \\  
 92962    &  9139163&18:56:22.13&+45:30:25.3&8.33&F8\,V   &   2 &   11 &  $-30.14\pm0.39$&  ... & ... \\  
\bf 92983 & 10454113&18:56:36.62&+47:39:23.1&8.62&F9\,V   &   1 &  ... &  $-24.14\pm0.22$&  ... & ... \\  
 93108A   & 10124866&18:58:03.46&+47:11:29.9&7.58&G0\,V   &  27 & 3606 &  $-29.81\pm0.12$& 59.4 & 0.00\\  
 93108B   & 10124866&18:58:03.46&+47:11:29.9&....&.....   &  28 & 3606 &  $-32.26\pm0.15$& 87.9 & 0.00\\  
 93236    &  7944142&18:59:28.70&+43:43:59.4&7.82&K1\,III &   2 &    7 &   $-1.00\pm0.42$&  ... & ... \\  
\bf 93556 & 11018874&19:03:19.20&+48:30:35.8&8.83&F6\,V   &   2 &   17 &    $0.65\pm0.61$&  ... & ... \\  
 93594    &  4818496&19:03:39.26&+39:54:39.4&8.14&A1\,V\footnote{\citet{Grenier1999}, 
 $^b$ \citet{Greenwich1935},
 $^c$ \citet{Bidelman1985}, 
 $^d$ \citet{Sato1990}, 
 $^e$ \citet{Wilson1962},
 $^f$ \citet{Thompson1978},
 $^g$ \citet{Gray2003}. \\
 $^h$ The individual radial velocities of the components are listed in Table 1.\\
 $^i$ The barycentric velocity of the system from Table 3.}
                                                          &   1 &  ... &   $14.28\pm7.82$&   ... & ... \\ 
 93657    &  5774694&19:04:16.37&+41:00:11.4&8.31&G1\,V   &  25 & 5479 &  $-17.76\pm0.13$&  40.1 & 0.01\\ 
\bf 93703 &  8547390&19:04:50.64&+44:41:28.9&8.38&K0\,III &   1 &  ... &  $-41.66\pm0.12$&   ... & ... \\ 
\bf 93898 &  8677933&19:07:12.19&+44:50:30.2&8.87&G0\,IV  &   1 &  ... &  $-14.41\pm0.54$&   ... & ... \\ 
 93951    & 11498538&19:07:46.85&+49:29:07.3&7.34&F4\,V   &   1 &  ... &  $-10.35\pm0.43$&   ... & ... \\ 
\bf 94071 &  3733735&19:09:01.92&+38:53:59.6&8.37&F4\,V   &   2 & 2874 &  $  3.10\pm0.44$&   ... & ... \\ 
 94112    &  3632418&19:09:26.83&+38:42:50.5&8.22&F8\,V   &   1 &  ... &  $-19.72\pm0.15$&   ... & SB1?\\ 
 94239    & 12453925&19:11:00.70&+51:21:43.8&8.34&F5\,V   &   1 &  ... &    $4.15\pm2.20$&   ... & ... \\ 
\bf 94343 &  6432054&19:12:09.53&+41:50:15.3&8.21&F0$^b$  &   1 &  ... &  $ 19.58\pm5.30$&   ... & ... \\ 
\bf 94472 & 12253106&19:13:38.78&+50:54:30.9&8.41&Am$^c$  &   1 &  ... &  $  5.39\pm1.39$&   ... & ... \\ 
 94675    & 10068307&19:15:54.70&+47:03:40.5&8.18&F8\,V   &   1 &  ... &  $-15.22\pm0.19$&   ... & ... \\ 
\bf 94798 & 11708170&19:17:18.50&+49:51:00.8&7.21&F1\,V   &   2 & 3272 &  $ -9.72\pm0.47$&   ... & ... \\ 
\bf 94922 & 11709006&19:18:56.54&+49:51:35.2&8.78&G1\,V   &   3 & 2797 &  $ -0.09\pm0.27$&   ... & ... \\ 
\bf 94924 &  4150611&19:18:58.20&+39:16:01.4&7.90&A5\,V$^d$&  3 &    5 &         ...$^h$ ~~~~~~&  ... &  SB2\\
 94931    &  6278762&19:19:00.55&+41:38:04.6&8.72&K0\,V$^e$&  23& 9189 & $-121.52\pm0.14$&  22.2 & 0.33\\
 95115    &  3641446&19:20:59.16&+38:47:23.8&8.37&F1\,V   &   1 &  ... &         ...$^h$ ~~~~~~&  ... &  SB2\\ 
\bf 95274 & 10532461&19:23:01.94&+47:42:52.5&8.85&G8\,V   &   3 & 2799 &  $  6.53\pm0.21$& ... & ... \\ 
\bf 95438 &  7820638&19:24:48.70&+43:31:18.5&7.95&G9\,III &   1 &  ... &  $-19.53\pm0.21$& ... & ... \\ 
\bf 95491 & 10010623&19:25:25.82&+46:57:29.1&8.40&F4\,V   &   2 &    4 &  $-22.30\pm0.30$& ... & ... \\ 
\bf 95548 &  6862114&19:26:05.47&+42:19:34.1&8.01&A2$^f$  &   1 &  ... &  $-11.11\pm4.50$& ... & ... \\ 
\bf 95549 &  3747220&19:26:05.93&+38:48:49.1&8.05&F1\,V   &   1 &  ... &  $ -7.72\pm1.05$& ... & ... \\ 
  95568   & 12258514&19:26:22.08&+50:59:14.1&8.08&G0      &   3 & 2799 &  $-19.61\pm0.15$& ... & ... \\ 
  95575   & 11506859&19:26:25.97&+49:27:55.0&7.90&K2.5\,V$^g$& 3& 5839 &         ...$^h$ ~~~~~~& ... & SB2\\  
\bf 95580 & 11189959&19:26:28.37&+48:52:14.9&8.21&A0$^b$  &   1 &  ... &  $ -8.14\pm6.23$& ... & ... \\ 
\bf 95661 & 11402951&19:27:32.81&+49:15:23.5&8.13&F0\,V   &   1 &  ... &  $-25.88\pm3.39$& ... & ... \\ 
  95876   &  8561664&19:29:56.69&+44:41:22.9&7.78&F3\,V   &   2 & 3308 &    $4.97\pm0.32$& ... & ... \\ 
\bf 96010 &  3347643&19:31:16.22&+38:24:02.9&8.02&A2$^b$  &   2 &    5 &  $-10.01\pm3.71$& ... & ... \\ 
  96062   & 11031993&19:31:56.02&+48:35:34.2&8.47&F5\,V   &  74 & 9481 &  $-57.74\pm0.17$& 70.3 & SB1?\\ 
  96528   &  5371516&19:37:27.70&+40:35:19.9&8.37&F6\,IV  &   1 &  ... &   $-5.56\pm0.25$& ... & ... \\ 
\bf 96561 &  7898839&19:37:51.72&+43:37:31.6&8.86&K0\,V   &   2 & 2879 &  $  4.26\pm0.32$& ... & ... \\ 
\bf 96775 &  4574610&19:40:15.43&+39:40:58.4&8.83&F6\,V   &   2 &    8 &  $-44.53\pm0.56$& ... & ... \\ 
\bf 97071 & 11253226&19:43:39.62&+48:55:44.2&8.44&F5\,IV  &   2 &    6 &  $  9.89\pm0.17$& ... & ... \\ 
\bf 97236 &  4484238&19:45:44.59&+39:30:13.2&8.56&F9\,V   &   2 & 2771 &  $-14.86\pm0.65$& ... & ... \\ 
\bf 97316 & 12317678&19:46:37.73&+51:01:13.5&8.74&F6\,V   &   1 &  ... &  $-26.39\pm0.19$& ... & ... \\ 
  97321   &  8379927&19:46:41.30&+44:20:54.7&6.96&F8\,V   &  44 & 5890 &    $1.06\pm0.07^i$& ...&  SB2\\
\bf 97341 & 11255615&19:47:00.98&+48:55:55.2&8.81&F7\,IV  &   2 &    4 &  $-17.29\pm0.13$& ... & ... \\ 
  97706   &  5557932&19:51:24.77&+40:44:07.4&8.14&G9\,V   &   3 & 2760 &  $-24.03\pm1.39$& ... & ... \\ 
\bf 97992 & 10162436&19:54:54.17&+47:08:43.0&8.61&F8\,V   &   1 &  ... &  $-54.80\pm0.17$& ... & ... \\ 
\hline\noalign{\smallskip}
\end{tabular} 
\end{center} 
\end{minipage}
\end{table*}

\section{Radial Velocity}

For stars observed at OAC, the radial velocity was derived with the cross-correlation method through the
{\sf fxcorr} task in IRAF. For stars of the spectral type F, G, and K we used Arcturus (K1.5~III, $RV 
= -5.30$ km\,s$^{-1}$ by \citet{udry1999}) or 54~Aql (F8~V, $RV = -0.20$ km\,s$^{-1}$ by \citet{evans1967}) 
as the templates. For the early-type stars, we used HR~1389 (A2~V-IV, $RV=38.97$ km\,s$^{-1}$ by \citet{fekel1999}).

The weighted mean radial velocities were calculated by averaging the measurements from all echelle orders, 
adopting the instrumental weight $W_i = \sigma _i^2$ for each $i$-th order. The values of the $\sigma _i^2$ errors
were computed by the {\sf fxcor} task in IRAF taking into account the height of the fitted peak and the 
antisymmetric noise \citep[see][]{tonry79}. The uncertainties in the weighted means of $RV$ were computed on 
the basis of $\sigma _i^2$ in each echelle order as described, e.g., by \citet{topping72}. 

For stars for which we acquired only two or three spectrograms, we computed both the external and the 
internal error of the weighted mean \citep[see][]{topping72} and adopted the higher of the two as the representative  
uncertainty.

In order to estimate any possible systematic error of the $RV$ derived from the spectrograms acquired at OAC, we
performed a self-consistency check of the $RV$ measured for Arcturus, one of our $RV$ standards observed for 1.5 years 
covering the time-span of our observations. We derived $RV$ of Arcturus using the first spectrum
acquired for this star as the template. As a result, we found that the derived values are self-consistent with 
the rms precision of better than 0.3 km\,s$^{-1}.$ 
An analogous analysis preformed by \citet{Stefanik99} shows that the $RV$ derived from the CfA observations are 
accurate to better than 0.2 km\,s$^{-1}.$

The spectra acquired at CfA span a single echelle order centered on 5187\,\AA\, that includes the Mg~b features. To derive the 
stellar parameters and radial velocities, the observed spectra were cross-correlated against a library of synthetic 
spectra calculated by John Laird using Kurucz model atmospheres and a line list developed by Jon Morse. The details
of the analysis are documented in \citet{latham2002}, along with a description of the procedures used to identify 
spectroscopic binaries and derive orbital solutions.

\begin{table}
\begin{center}
\caption{HIP\,97321: Orbital Solution}
\label{hip97321-fig}
\begin{tabular}{ll}
\hline\hline\noalign{\smallskip}
$P$                  = 1730.3 $\pm$ 5.8 d\\
$\gamma$             = 1.063 $\pm$ 0.070 km\,s$^{-1}$\\
$K_{\rm A}$          = 9.13 $\pm$ 0.11 km\,s$^{-1}$\\
$K_{\rm B}$          = 11.87 $\pm$ 0.30 km\,s$^{-1}$\\
$e$                  = 0.231 $\pm$ 0.010 \\
$\omega _1$          = 85.5$\rm ^o$ $\pm$ 2.9$\rm ^o$ \\
$T$                  = HJD 2450219 $\pm$ 14 \\
$a_{\rm A} \sin i$   = 211.4 $\pm$ 2.3 $\times 10 ^6$ km\\
$a_{\rm B} \sin i$   = 274.9 $\pm$ 6.6 $\times 10 ^6$ km\\
$M_{\rm A} \sin^3 i$ = 0.865 $\pm$ 0.046 $\rm M_{\odot}$\\
$M_{\rm B} \sin^3 i$ = 0.666 $\pm$ 0.024 $\rm M_{\odot}$\\ 
$q$                  = 0.769 $\pm$ 0.020 \\
\hline
\end{tabular}
\end{center}
\end{table}

\begin{figure}
\includegraphics[width=8.5cm]{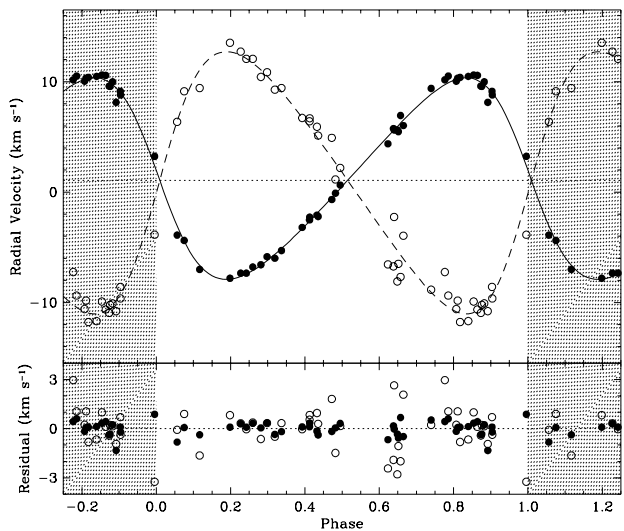}
\caption[]{\textit{Top:} The $RV$ light curve of the primary (dots) and the secondary (circles) component 
of the SB2 system HIP\,97321. \textit{Bottom:} The residuals of $RV$ of the components left after subtracting the 
orbital motion.}
\label{HIP97321}
\end{figure}

{\bf The individual radial velocity measurements for our program stars are given in Table~\ref{rv-individual}. 
This Table is available only in the electronic form. Only a sample, containing the first four rows, and the last row, 
is presented.} In the first and the second column of this Table, we give the HIP number from the Hipparcos Catalogue 
\citep{esa97} and the KIC number from Kepler Input Catalog. Then, in the following columns, the right ascension and 
declination of the star for the epoch 2000, the Kepler magnitude of the star, the Heliocentric Julian Day of the 
middle of the exposure, the radial velocity and its standard error, and the name of the instrument used for acquiring the 
spectrum are quoted. In the two last columns, we list the exposure times and the signal-to-noise ratios for the spectrograms 
acquired with FRESCO; at the remaining observing sites, the exposure time was set to 1200\,s, and the typical S/N was 
15-20.

In Table 1, we do not list the errors of $RV$ of the components of the double-lined spectroscopic 
binaries HIP\,95575 and HIP\,97321 for which the $RV$ were computed using the TODCOR two-dimensional 
correlation technique \citep{zucker94} because TODCOR does not provide uncertainties for individual observations. 
Instead, in Fig.~\ref{HIP97321} we provide the rms velocity residuals from the orbital solution for the primary 
and secondary component of HIP\,97321, which give a good overview of the velocity errors.

In Table~\ref{param}, we list the HIP and KIC numbers of all the observed targets, 
their equatorial coordinates for the epoch 2000, the \Kepler magnitude, the spectral type derived 
in this paper or adopted from the literature, the number of the acquired spectrograms, the time-span 
of the observations, the mean radial velocity and its standard deviation.
Then, we list the sum of the residuals divided by the internal error estimate squared, $\chi ^2$, and the probability 
that a star with constant velocity will have $\chi ^2$ values larger than the observed one, $\rm P(\chi ^2)$. 
For the spectroscopic binaries, in the last column we give the classification to the type of the spectroscopic 
variability. For 25 stars whose HIP numbers in Table~\ref{param} are typed bold face, the radial velocity given in 
this paper has been measured for the first time.

We note that the mean radial velocities given in Table~\ref{param} were calculated subtracting 0.5 km\,s$^{-1}$ from 
each measurement of $RV$ derived from the spectra acquired at OAC in order to put all the measurements on the native CfA system, 
as explained in \citet{molenda07}. To put these values on an absolute system, 0.14 km\,s$^{-1}$ should be added to the CfA 
native velocities \citep[see][noting that the sign of the correction in that paper is in error.]{Stefanik99} 

\subsection{Stars variable in $RV$}

Our sample contains four double-lined spectroscopic binaries (SB2): HIP\,94924, HIP\,95115, HIP\,95575, and HIP\,97321. 
Three of them, namely, HIP\,94924, HIP\,95115, and HIP\,97321, have been discovered in this paper; HIP\,95575 was 
discovered to be SB2 by \citet{Tokovinin1991} who also provides the orbital solution for this system. That star has 
been then observed with speckle interferometry by \citet{Mason1999} but not resolved.

For one of the new SB2 stars, HIP\,97321, we acquired 44 spectrograms that allowed us to compute the orbital 
solution given in Table~\ref{hip97321-fig}. In Fig.~\ref{HIP97321}, we plot the $RV$ light curve of the 
primary and the secondary component of this system. As can be seen in this figure, typical uncertainties of the 
$RV$ measurements of the primary and the secondary component of HIP\,97321 are $\pm 1$ and $\pm 2$ km\,s$^{-1}$, respectively.

For the three other SB2 systems, HIP\,94924, HIP\,95115, and HIP\,95575, we acquired only few spectrograms. Therefore, more 
data are needed to compute the orbital solutions of these stars.

For HIP\,94112 ($\langle RV \rangle = -19.72\pm0.15$ km\,s$^{-1}$) and HIP\,96062 ($\langle RV \rangle = -57.74\pm0.19$ km\,s$^{-1}$) 
the radial velocities measured by us differ by more than $3\sigma$ from the values reported in the literature which 
amount to $-28$ km\,s$^{-1}$ for HIP\,94112 \citep{Moore1950} and $-$46.6 km\,s$^{-1}$ for HIP\,96062 \citep{Fouts1986}.  
However, our determinations agree with $-18.9\pm0.3$ km\,s$^{-1}$ and $-57.8\pm0.2$ km\,s$^{-1}$ given for these two stars 
by \citet{nordstrom2004} and \citet{latham2002}, respectively. We classify HIP\,94112 and HIP\,96062 as suspected 
single-lined spectroscopic binaries.

Finally, we acquired 28 spectrograms spanning 3606 days of the speckle binary HIP\,93108 which is a close north-south 
pair with nominal separation $1^{\prime\prime}.7$. We found that the mean velocities of the components differ by about 
2.6~km\,s$^{-1}$ but we did not detect any obvious acceleration of either star. 

We note that none of the double-lined and suspected single-lined binaries discovered in this paper is known to be 
variable photometrically.

\section{Atmospheric Parameters}

\subsection{From Comparison with Standard Stars}

For 33 F, G, and K stars observed at OAC, we determined the $T_{\rm eff}$, $\log g$ and 
$\rm [Fe/H]$ using the ROTFIT code \citep[see][2006]{Frasca2003} as described in detail by \citet{molenda07}. 
The method, which has been originally developed by \citet{Katz1998} and \citet{Soubiran1998}, consists in comparing 
the spectra of program stars with a library of spectra of reference stars. The atmospheric parameters of the program 
stars are computed as the weighted means of the astrophysical parameters of the reference stars which best 
reproduce the target spectrum. For measuring the similarity of spectra, the value of $\chi ^2$ is used.
Then, the weighted standard error ($\sigma$) is computed per each order and the means of each order are averaged using
$(\sigma ^{-2} \chi ^{-2}f)$ as a weight. Here, the factor $\chi ^{-2}$ accounts for differences between orders due to 
different S/N and the goodness of the fit, and the factor $f$ is proportional to the total absorption of lines in 
each individual order. The $f$ factor approximately corrects for the different amount of information contained 
in the blue orders and the red orders with few lines and much continuum, and gives also more weight to orders containing strong
and broad lines. As discussed, e.g., by \citet{Frasca2006}, this method allows a fast determination 
of $T_{\rm eff}$, $\log g$, and $\rm [Fe/H]$ of stars of the spectral type F, G, and K, and
it can be successfully applied even to spectrograms of low signal-to-noise ratio and moderate resolution.

We used two separate sets of reference spectra to perform independent determinations of
$T_{\rm eff}$, $\log g$, and $\rm [Fe/H]$ of our targets. The first set consisted of ELODIE archive spectra
\citep{Prugniel2001} of 246 slowly-rotating stars, the second, of 122 slowly rotating stars 
observed at OAC with FRESCO. The atmospheric parameters of the 246 stars from the ELODIE archive and of 
109 stars from the FRESCO archive have been given by \citet{molenda08}. The adopted atmospheric parameters 
of 13 new reference stars observed with FRESCO in 2009 are given in Table~\ref{13new}, the last column 
of which contains the references to the bottom of the table where we give the sources of the adopted values 
of $T_{\rm eff}$, $\log g$, and [Fe/H].

For computing the atmospheric parameters of stars for which we have multiple exposures, we used                                
the spectrograms of the highest signal-to noise ratio. The values of $T_{\rm eff}$, $\log g$, $\rm [Fe/H]$, and 
their uncertainties derived by means of the ROTFIT method are given in Table~\ref{teff}. 

In the sixth column of Table~\ref{param}, we list also the MK types of the program stars which we inferred by 
adopting the spectral type and the luminosity class of the reference stars which occurred most frequently. 
For stars for which we did not perform the spectral classification, we give the spectral classification from the literature.

\begin{figure*}
\includegraphics[width=14cm,angle=0]{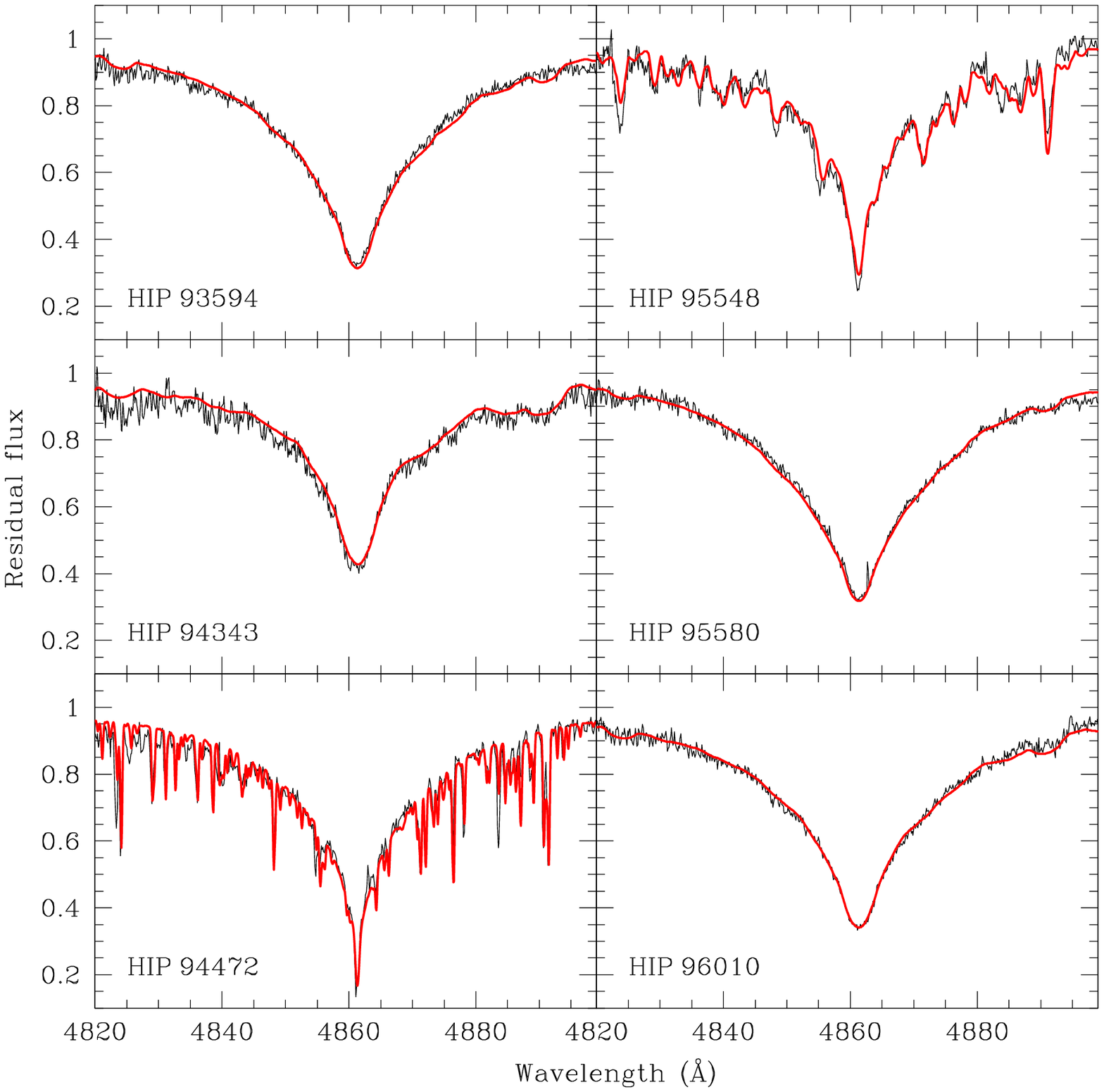}
\caption[]{The results of fitting of the H$\beta$ line for the early-type stars observed at OAC.}
\label{hbeta}
\end{figure*}

\begin{table}
\begin{center}
\caption{The astrophysical parameters of 13 new reference stars observed at OAC.}
\label{13new}
\begin{tabular}{r r c r c c}
\hline
\hline
    HD  & $T_{\rm eff}$ & $\log g$ & [Fe/H] & ref.\\  
        & (K)           & (dex)    & (dex)\\
\hline\noalign{\smallskip}
  84441 & 5310 & 2.04 & $-$0.28 & 1\\
 157881 & 4180 & 4.70 & $-$0.20 & 2\\
 169931 & 3215 & 0.50 &    0.00 & 3\\
 186427 & 5760 & 4.40 &    0.06 & 2\\
 190007 & 4650 & 4.26 &    0.15 & 4\\
 191026 & 5150 & 3.49 & $-$0.10 & 2\\
 196755 & 5510 & 3.60 & $-$0.09 & 2\\
 198149 & 4950 & 3.41 & $-$0.32 & 2\\
 201091 & 4500 & 4.56 & $-$0.43 & 2\\
 201092 & 4120 & 4.40 & $-$0.63 & 2\\
 204613 & 5650 & 3.80 & $-$0.35 & 2\\
 212943 & 4582 & 2.75 & $-$0.30 & 2\\
 222404 & 4810 & 3.00 &    0.04 & 2\\
\hline\noalign{\smallskip}
\end{tabular}
\begin{list}{}
\item [1] \citet{McWILLIAM1990}, [2] \citet{Cayrel2001}, [3] \citet{Prugniel2007}, [4] \citet{Soubiran2008}
\end{list}
\end{center}
\end{table}

\begin{table*}
\begin{minipage}{17.7cm}
\caption{Atmospheric parameters for F, G, and K stars determined with the use of the ELODIE and FRESCO reference stars.}
\label{teff} 
\begin{center}                
\begin{tabular}{lrrrrrrrr}     
\hline                  
\hline            
HIP     &~~~~T$_{\rm eff}$& $\log g$     & [Fe/H]   &~~~~T$_{\rm eff}$    & $\log g$  & [Fe/H]       & $v\,\sin i$ \footnote{Derived by means of the ROTFIT code.}~ & $v\,\sin i$ \footnote{Derived by means of the FWHM method.}~\\
        &~~\tiny ELODIE&\tiny ELODIE&\tiny ELODIE   &~~\tiny FRESCO&\tiny FRESCO&\tiny FRESCO        & (km s$^{-1}$) & (km s$^{-1}$)\\
\hline                                     
91990 & 6446$\pm$195 & 4.07$\pm$0.07 &             $-$0.02$\pm$0.09 & 	6472$\pm$~98 & 4.02$\pm$0.08 &             $-$0.11$\pm$0.06 & 53$\pm$2 	 & 52$\pm$2 \\
92962 & 6078$\pm$108 & 4.11$\pm$0.06 &             $-$0.04$\pm$0.08 & 	6056$\pm$~75 & 3.99$\pm$0.06 &             $-$0.03$\pm$0.06 &  $<5$      & $<5$     \\
92983 & 5884$\pm$135 & 4.13$\pm$0.07 &             $-$0.12$\pm$0.10 & 	6030$\pm$102 & 4.20$\pm$0.10 &             $-$0.22$\pm$0.06 &  $<5$      & $<5$     \\
93108\footnote{The atmospheric parameters and the $v\sin i$ are derived from the composite spectrum.}
 & \it 5749$\pm$ \it 65 & \it 4.31$\pm$0.06 & \it  $-$0.28$\pm$0.06 &\it 5980$\pm$108 &\it 4.17$\pm$0.10 & \it     $-$0.21$\pm$0.07 &  $<5$      & $<5$     \\
93236 & 4541$\pm$~67 & 2.44$\pm$0.16 & \hspace{6.5pt} 0.03$\pm$0.06 & 	4667$\pm$100 & 2.53$\pm$0.22 & \hspace{6.5pt} 0.00$\pm$0.07 &  $<5$      & $<5$     \\
93556 & 6326$\pm$172 & 4.02$\pm$0.09 &             $-$0.09$\pm$0.11 & 	6151$\pm$~99 & 4.07$\pm$0.10 &             $-$0.23$\pm$0.06 & 56$\pm$2 	 & 56$\pm$4 \\
93657 & 5798$\pm$~68 & 4.30$\pm$0.05 & \hspace{6.5pt} 0.01$\pm$0.06 & 	5924$\pm$116 & 4.16$\pm$0.12 & \hspace{6.5pt} 0.05$\pm$0.06 &  $<5$      & $<5$     \\
93703 & 4676$\pm$~70 & 2.42$\pm$0.16 &             $-$0.03$\pm$0.07 & 	4790$\pm$~49 & 2.73$\pm$0.10 & \hspace{6.5pt} 0.04$\pm$0.06 &  $<5$      & $<5$     \\
93898 & 5893$\pm$137 & 3.20$\pm$0.27 & \hspace{6.5pt} 0.02$\pm$0.09 & 	6032$\pm$100 & 3.25$\pm$0.21 & \hspace{6.5pt} 0.06$\pm$0.05 & 50$\pm$2	 & 47$\pm$3 \\
93951 & 6574$\pm$~84 & 4.06$\pm$0.06 &             $-$0.02$\pm$0.05 & 	6481$\pm$148 & 4.01$\pm$0.12 &             $-$0.04$\pm$0.06 & 39$\pm$2 	 & 40$\pm$2 \\
94071 & 6539$\pm$120 & 4.03$\pm$0.06 &             $-$0.06$\pm$0.08 & 	6365$\pm$~87 & 4.10$\pm$0.13 &             $-$0.18$\pm$0.06 & 17$\pm$2 	 & 17$\pm$2 \\
94112 & 6063$\pm$126 & 4.04$\pm$0.07 &             $-$0.23$\pm$0.09 & 	6145$\pm$~65 & 4.15$\pm$0.10 &             $-$0.15$\pm$0.06 &  $<5$      & $<5$     \\
94239 & 6488$\pm$201 & 4.10$\pm$0.09 &             $-$0.03$\pm$0.14 & 	6417$\pm$121 & 4.03$\pm$0.20 &             $-$0.12$\pm$0.06 & 82$\pm$4	 & 80$\pm$8 \\
94675 & 6059$\pm$166 & 4.02$\pm$0.07 &             $-$0.30$\pm$0.07 & 	6022$\pm$~65 & 4.07$\pm$0.06 &             $-$0.26$\pm$0.06 &  $<5$      & $<5$     \\
94798 & 6851$\pm$134 & 4.12$\pm$0.09 &             $-$0.02$\pm$0.08 &   ...~~~~~     & ...~~~~~      &                     ...~~~~~ & 39$\pm$4   & ...~  \\
94922 & 5861$\pm$~92 & 4.30$\pm$0.09 &             $-$0.06$\pm$0.07 & 	6019$\pm$~84 & 4.20$\pm$0.11 & \hspace{6.5pt} 0.00$\pm$0.06 &  9$\pm$2   & 10$\pm$2 \\
95115$^c$ & \it 6745$\pm$140 & \it 4.11$\pm$0.07 & \it \hspace{6.5pt}  0.02$\pm$0.06 & ...~~~~~ & ...~~~~~       &  ...~~~~     & \it 66$\pm$3   & ...~  \\
95274 & 5432$\pm$~89 & 4.38$\pm$0.18 &             $-$0.07$\pm$0.07 & 	5559$\pm$180 & 4.22$\pm$0.12 &             $-$0.03$\pm$0.07 &  $<5$      & $<5$     \\
95438 & 4837$\pm$~66 & 2.91$\pm$0.17 &             $-$0.05$\pm$0.06 & 	4784$\pm$~56 & 2.79$\pm$0.08 &             $-$0.05$\pm$0.09 &  $<5$      & $<5$     \\
95491 & 6499$\pm$155 & 4.08$\pm$0.06 &             $-$0.02$\pm$0.07 & 	6277$\pm$138 & 4.05$\pm$0.07 &             $-$0.14$\pm$0.06 & 35$\pm$5	 & 35$\pm$3 \\
95549 & 6833$\pm$126 & 4.19$\pm$0.06 &                0.00$\pm$0.06 &   ...~~~~~     &  ...~~~~~     &                    ...~~~~~  & 77$\pm$5   & ...~~\\
95661 & 7026$\pm$202 & 4.13$\pm$0.10 &             $-$0.03$\pm$0.17 &   ...~~~~~     &  ...~~~~~     &                    ...~~~~~  & 91$\pm$6   & ...~~\\
95876 & 6149$\pm$~72 & 4.00$\pm$0.04 &             $-$0.62$\pm$0.06 & 	6220$\pm$121 & 4.03$\pm$0.11 &             $-$0.42$\pm$0.17 & 12$\pm$6   & 12$\pm$4 \\
96062 & 6430$\pm$168 & 4.12$\pm$0.09 & \hspace{6.5pt} 0.05$\pm$0.08 & 	6177$\pm$~75 & 4.02$\pm$0.16 &             $-$0.04$\pm$0.06 & 64$\pm$2   & 65$\pm$3 \\
96528 & 6117$\pm$103 & 4.07$\pm$0.06 & \hspace{6.5pt} 0.02$\pm$0.06 & 	6136$\pm$~78 & 4.06$\pm$0.06 &             $-$0.05$\pm$0.05 & 12$\pm$2   & 13$\pm$2 \\
96561 & 5266$\pm$~67 & 4.42$\pm$0.19 & \hspace{6.5pt} 0.01$\pm$0.06 & 	5163$\pm$~74 & 4.13$\pm$0.18 &             $-$0.07$\pm$0.06 &  $<5$      & $<5$     \\
96775 & 6341$\pm$121 & 4.06$\pm$0.06 &                0.00$\pm$0.07 & 	6298$\pm$110 & 4.01$\pm$0.08 &             $-$0.10$\pm$0.05 & 32$\pm$3   & 33$\pm$2 \\
97071 & 6369$\pm$130 & 3.98$\pm$0.06 &             $-$0.19$\pm$0.07 & 	6283$\pm$~98 & 4.00$\pm$0.10 &             $-$0.25$\pm$0.05 & 12$\pm$2   & 12$\pm$2 \\
97236 & 5998$\pm$104 & 4.14$\pm$0.06 &             $-$0.03$\pm$0.08 & 	6045$\pm$~64 & 4.02$\pm$0.10 &             $-$0.04$\pm$0.06 &  $<5$      & $<5$     \\
97316 & 6327$\pm$137 & 3.99$\pm$0.07 &             $-$0.32$\pm$0.08 & 	6413$\pm$~81 & 4.03$\pm$0.06 &             $-$0.29$\pm$0.07 &  $<5$      & $<5$     \\         
97341 & 6158$\pm$125 & 3.97$\pm$0.07 &             $-$0.32$\pm$0.07 & 	6078$\pm$~71 & 4.07$\pm$0.07 &             $-$0.37$\pm$0.06 &  $<5$      & $<5$     \\
97992 & 5995$\pm$190 & 4.02$\pm$0.07 &             $-$0.28$\pm$0.08 & 	6074$\pm$120 & 4.09$\pm$0.07 &             $-$0.15$\pm$0.06 &  $<5$      & $<5$     \\
\hline\noalign{\smallskip}
\end{tabular}
\end{center}
\end{minipage}
\end{table*}

\begin{table*}
\begin{minipage}{17.7cm}
\caption{Atmospheric parameters, 
chemical abundances, and $v\sin i$s for early-type stars from model atmospheres. The abundance
values are in solar units \citep{asplund05}. The uncertainties of the derived values are 
given in the brackets.}
\label{model}
\begin{center}
\begin{tabular}{lccrrrrrrrrrrr}
\hline
\hline
HIP     &T$_{\rm eff}$& $\log g$ & 
$\left[ Fe \right]$ & 
$\left[ C  \right]$ &
$\left[ O  \right]$ &
$\left[ Na \right]$ &
$\left[ Mg \right]$ &
$\left[ Si \right]$ &
$\left[ S  \right]$ &
$\left[ Ti \right]$ &
$\left[ Cr \right]$ &
$\left[ Ni \right]$ &
$v\sin i$ \\ 
\hline\noalign{\smallskip}
93594  & 8400 & 3.8 & $-$0.4 &...&...&...& $-$0.1 &...&...&...&...&...& 150 \\ \vspace{0.2cm}
       & (300)&(0.3)&   (0.1)&...&...&...&   (0.1)&...&...&...&...&...& (28)\\ 

94343  & 7100 & 3.5 &    0.0 &...&...&...& $-$0.2 &...&...&...&...&...& 200\\ \vspace{0.2cm}
       & (200)&(0.3)&  (0.2) &...&...&...&  (0.2) &...&...&...&...&...& (10)\\ 

94472  & 7700 & 4.0 &    0.6 & 0.4 & 0.6 & 0.8 & 0.2 &    0.3 & 0.6 & 0.5 & 1.0 & 1.0 & 13 \\ \vspace{0.2cm}
       & (200)&(0.2)&  (0.1) &(0.2)&(0.3)&(0.3)&(0.3)&   (0.3)&(0.3)&(0.2)&(0.3)&(0.2)&(2)\\

95548  & 7200 & 4.0 &    0.5 & 0.7 & 0.3 & 0.4 & 0.1 & $-$0.3 & 0.4 & 0.7 & 0.1 & 0.7 & 52\\ \vspace{0.2cm}
       & (200)&(0.2)&  (0.2) &(0.3)&(0.3)&(0.3)&(0.3)& (0.3)  &(0.2)&(0.2)&(0.1)&(0.2)& (4)\\

95580  & 9200 & 4.1 &    0.0 &...&...&...&    0.0 &...&...&...&...&...& 139\\ \vspace{0.2cm}
       & (300)&(0.2)&   (0.1)&...&...&...&  (0.1) &...&...&...&...&...&(13)\\

96010  & 8100 & 4.3 &    0.0 &...&...&...&    0.0 &...&...&...&...&...& 170\\ \vspace{0.2cm}
       &(200) &(0.2)&   (0.1)&...&...&...&  (0.1) &...&...&...&...&...& (10)\\
\hline\noalign{\smallskip}
\end{tabular}
\end{center}
\end{minipage} 
\end{table*}

\begin{figure}
\includegraphics[width=8.5cm,angle=0]{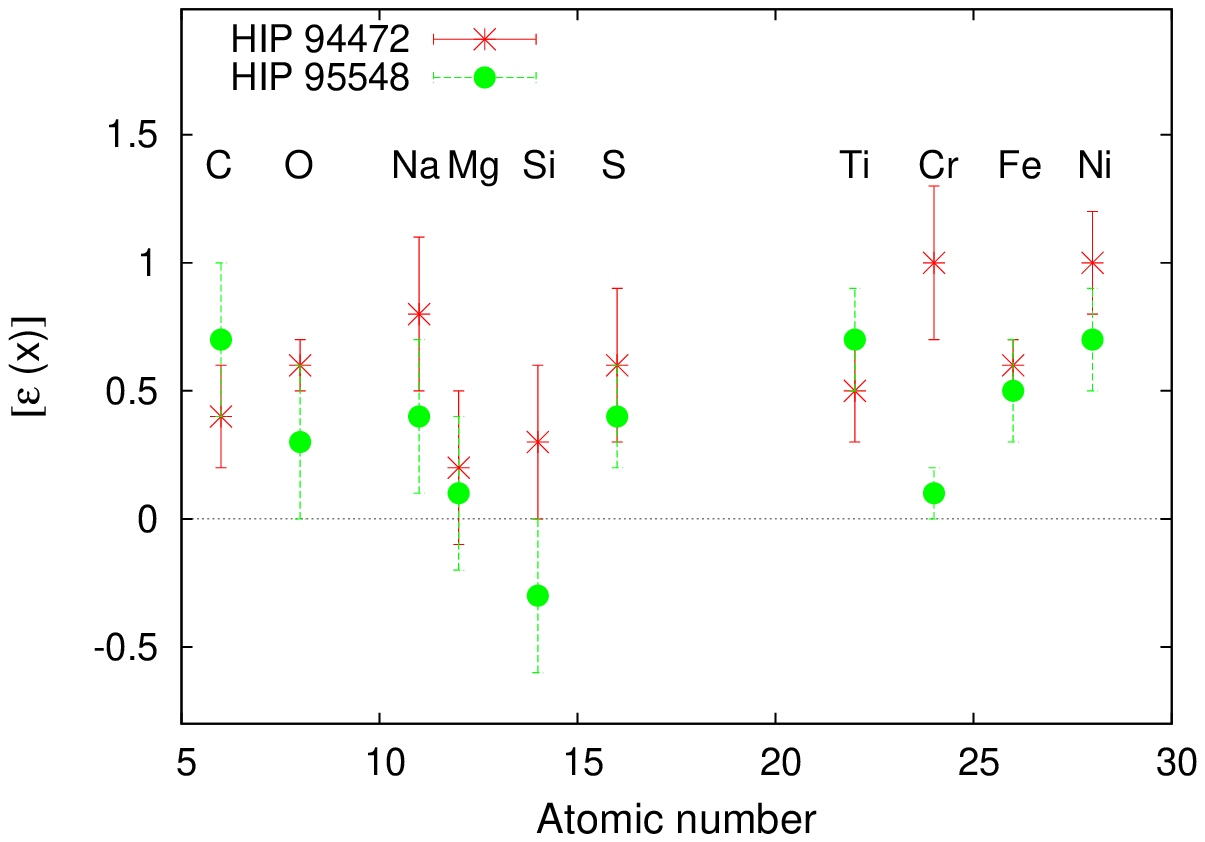}
\caption[]{The abundance patterns derived for HIP\,94472 and HIP\,95548 reported in Table~\ref{model}.
For the sake of clarity we excluded from the plot all the chemical elements
with Z\,$>$\,28 (i.e. nickel).}
\label{pattern}
\end{figure}

\subsection{From Model Atmospheres}
\subsubsection{Stars observed at OAC}

The atmospheric parameters of the hot stars observed at OAC, i.e., HIP\,94343, HIP\,93594, HIP\,94472,  
HIP\,95548, HIP\,95580, and HIP\,96010, were determined by minimizing the difference between the observed and the 
synthetic H$\beta$ profiles, using the $\chi^2$ defined as
\begin{equation}
\chi^2 = \frac{1}{N} \sum \bigg(\frac{I_{\rm obs} - I_{\rm th}}{\delta I_{\rm obs}}\bigg)^2
\end{equation}
for the measure of the goodness-of-fit parameter, as described in \citet{catanzaro10}. For starting estimations 
of T$_{\rm eff}$ and $\log g$, we used the values from the Kepler Input Catalog. 

In Fig.~\ref{hbeta}, we show the results of fitting the synthetic H$\beta$ line, computed 
with SYNTHE \citep{kur81} on the basis of ATLAS9 \citep{kur93} atmosphere models,
to the observed H$\beta$ in the six stars. All the models have been evaluated for
solar Opacity Distribution Function and microturbulent velocity $\xi$\,=\,2~km~s$^{-1}$.

For the slowly rotating stars HIP\,94472 and HIP\,95548, we determined the photospheric 
abundances by computing synthetic spectra that reproduce the observed ones. We therefore, divided the 
measured spectrograms into several intervals, each 25 {\AA} wide, and derived the abundances in each
interval by performing a $\chi2$ minimization of the difference between the observed and
synthetic spectrum. We adopted lists of spectral lines and atomic parameters
from \citet{castelli04}, who updated the parameters listed originally by \citet{kur95}.

We computed the abundances relative to the solar standard values given by \citet{asplund05}. For 
each element, we calculated the uncertainty in the abundance to be the standard deviation of the mean 
obtained from individual determinations in each interval of the analyzed spectrum. For the elements 
whose lines occurred in one or two intervals only, the error in the abundance was evaluated by varying 
the effective temperature and gravity within their uncertainties, $[ T_{\rm eff}\,\pm\, \delta T_{\rm eff}]$ 
and $[\log g\,\pm\,\delta \log g]$, and computing the abundance for $T_{\rm eff}$
and $\log g$ values in these ranges.

For the other four fast rotators, the lines are too broad to attempt this kind of analysis and we 
derived only the iron and magnesium abundances from the equivalent widths of Fe{\sc ii} $\lambda 
\lambda$5018.44, 5316.615~{\AA} and from Mg{\sc ii} $\lambda$4481 {\AA}. The latter were converted into
abundances using WIDTH9 \citep{kur81} and the ATLAS9 atmospheric models.

In Table~\ref{model}, we list the derived values of $T_{\rm eff}$, $ \log g$, $v\sin i$, the abundances,
and the uncertainties of these quantities.

\subsubsection{Stars observed with the CfA Digital Speedometers}

The atmospheric parameters of stars observed with the CfA Digital Speedometers were derived using synthetic spectra computed by 
Jon Morse (see Sect.\ 3) and one-dimensional correlations to identify the template that gives the best 
match with the observed spectrum. The 7650 template spectra that we used covered a range of [4000, 7500] 
in $T_{\rm eff}$, [0, 70] in $v\sin i$, [2.5, 5.0] in $\log g$, and [$-1.5\, ,$ 0.5] in [Fe/H]. The template 
that we chose gave the highest peak correlation value averaged over all the observed spectra for each program 
star. 

For many of the stars observed with the CfA Digital Speedometers we acquired only a single spectrogram, and some of these 
exposures are not long enough to
reliably extract information about the metallicity because of degeneracy between $T_{\rm eff}$, 
[Fe/H], and $\log g$ in the spectrum. For this reason, we treated stars with a weak exposure (fewer 
than 100 counted photons per pixel, corresponding to a signal-to-noise ratio of about 20 per spectral 
resolution element) differently than stars with a long enough exposure or multiple exposures. For stars 
with only one weak exposure, we fixed the metallicity to the grid value closest to that determined by 
ELODIE and FRESCO (or in the literature if that data was not available), and used a cubic spline to 
interpolate to the peak values in $\log g$, $T_{\rm eff}$, and $v\sin i$. For stars with long enough 
or multiple exposures, we additionally interpolated to the peak [Fe/H] value. 
The values coming out from our analysis are given in Table~\ref{model-dl}.

Since the limited number of our observations does not allow us to evaluate statistically significant errors, 
we adopt the grid spacing, i.e., 125\,K in $T_{\rm eff}$, 2 km\,s$^{-1}$ in $v\sin i$, and 0.25 dex in $\log g$ 
and [Fe/H] as the error estimate.

\begin{figure}
\includegraphics[width=8.5cm,angle=0]{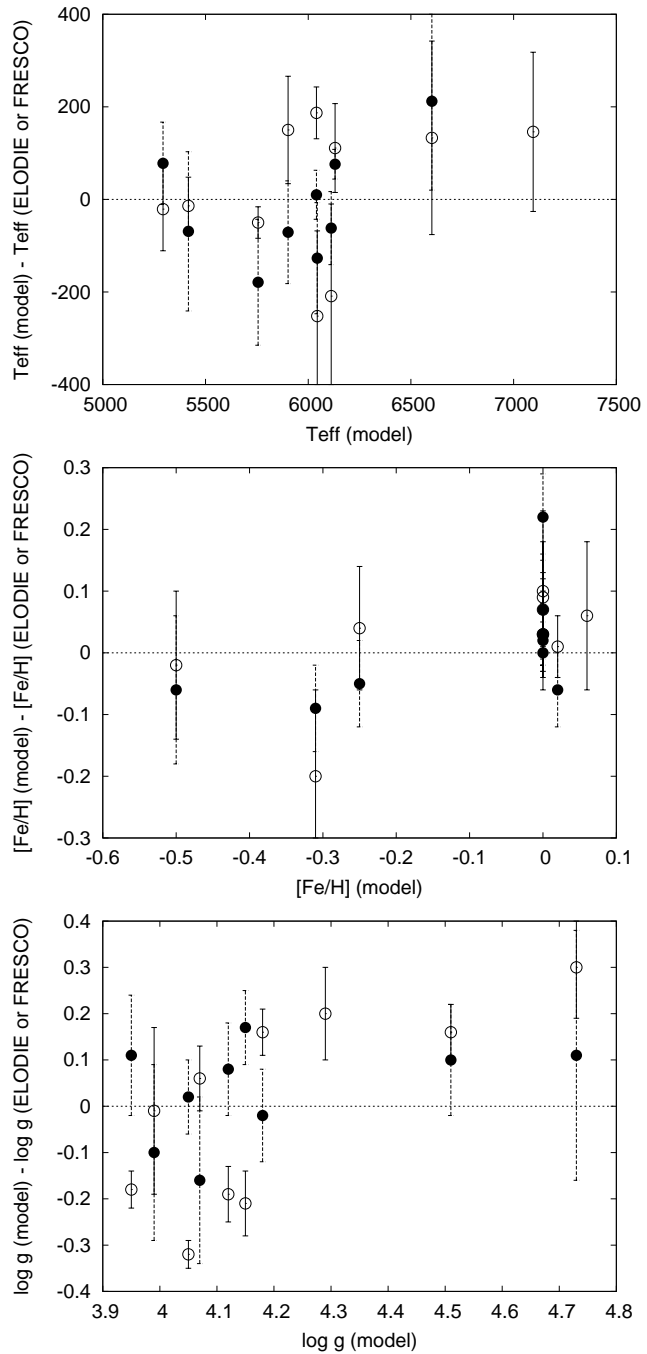}
\caption[]{The differences between 
$T_{\rm eff}$ ({\it top panel}), [Fe/H] ({\it middle panel}), and $\log g$
({\it bottom panel}) derived from the model atmospheres (stars observed with the CfA Digital Speedometers) and with the ROTFIT code
(stars observed at OAC.) {\it Filled circles:} results obtained with the FRESCO reference stars, {\it open circles:}
results obtained with the ELODIE reference stars. The zero line is indicated with the dotted lines.}
\label{model-elo-fre}
\end{figure}

In Fig.~\ref{model-elo-fre}, we show the differences between $T_{\rm eff}$, [Fe/H], and $\log g$ derived 
from the model atmospheres for stars observed with the CfA Digital Speedometers, and the respective parameters computed with 
the ROTFIT code when using the FRESCO or the ELODIE references stars. In most cases the values of $T_{\rm eff}$, [Fe/H], 
and $\log g$ agree to within $1\sigma$ error bars. We note, though, that the agreement is better for the FRESCO 
reference stars. This is not surprising, because this grid is composed of reference spectra taken with the same 
instrument as the target ones.

\begin{table}
\begin{minipage}{8cm}
\caption{The atmospheric parameters derived from the model atmospheres 
and the projected rotational velocity, $v\sin i$, for the program stars 
observed with the CfA Digital Speedometers.}
\label{model-dl}
\begin{center}
\begin{tabular}{lccrr}
\hline
\hline
HIP    & $T_{\rm eff}$ & $\log g $ & [Fe/H] & $v\sin i$ \\
       & (K)           & (dex)     & (dex)  & (km\,s$^{-1}$)\\
\hline\noalign{\smallskip}
93108A & 5916 & 4.48 &$ -0.27 $ & 0.6\\
93108B & 5888 & 4.55 &$ -0.23 $ & 2.0\\
93556  & 6112 & 4.07 &$ -0.31 $ &52.6\\
93657  & 5756 & 4.15 &$  0.02 $ & 4.6\\
94071  & 6602 & 4.18 &$  0.00 $ &17.4\\
94798  & 7095 & 4.29 &$  0.06 $ &36.8\\
94922  & 6040 & 4.12 &$  0.00 $ &10.3\\
94931  & 4963 & 4.36 &$ -0.53 $ & 1.1\\
95274  & 5417 & 4.05 &$  0.00 $ & 5.0\\
95568  & 5760 & 3.59 &$  0.00 $ & 5.6\\
95575\footnote{The atmospheric parameters and the $v\sin i$ are derived from the composite spectrum.}  
       & \it 4734 & \it 4.21 & \it $-$0.37&\it 13.5\\
95876  & 6044 & 3.99 &$ -0.50 $ &11.2\\
96062  & 6266 & 4.04 &$  0.00 $ &63.5\\
96561  & 5293 & 4.73 &$  0.00 $ & 3.0\\
97236  & 6131 & 3.95 &$  0.00 $ & 9.1\\
97321\,A\footnote{The stellar parameters are those that maximize the
weighted mean correlation coefficient in the TODCOR solution that uses fixed $\log g=4.5$ and [m/H]=0.}
 & 6350 & 4.50 &$  0.00 $ & 6.0\\ 
97321\,B$^b$ & 5500 & 4.50 &$  0.00 $ & 2.0\\
97706  & 5730 & 3.75 &$  0.00 $ &15.2\\
\hline
\end{tabular}
\end{center}
\end{minipage}
\end{table}

\begin{figure}
\includegraphics[width=8.5cm]{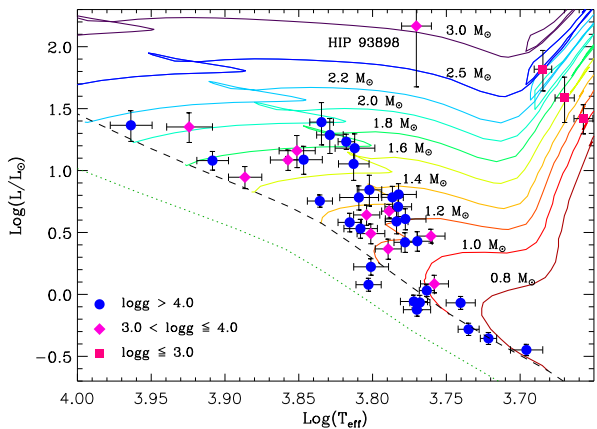}
\caption[]{The HR diagram for stars discussed in this paper. Different symbols are used to indicate stars of 
different $\log g$. The evolutionary tracks from \citet{Girardi2000} are plotted with solid lines; the ZAMS 
for $Z=0.019$ and $Z=0.001$ is plotted, respectively, with the dashed and dotted lines.}
\label{hrdiagram}
\end{figure}

We put the program stars on the Hertzsprung-Russel (HR) diagram in  Fig.~\ref{hrdiagram},
using different symbols for stars of $\log g \le 3.0$, $3.0 < \log g \le 4.0$, and $\log g 
> 4.0$, and using the Hipparcos parallaxes revised by \citet{vanLeeuwen07}, and $T_{\rm eff}$ from 
Tables~\ref{teff}, \ref{model} or \ref{model-dl} in this paper. The evolutionary tracks for the 
solar metallicity as well as the location of the ZAMS for $Z= 0.019$ (the solar metallicity)
and $Z=0.001$ are adopted from \citet{Girardi2000}. As can be seen in this figure, the location of 
the program stars in the HR diagram is consistent with the values of $\log g$ derived in this paper.

\section{Projected rotational velocity}

To derive $v\sin i$ of the F, G, and K type stars stars observed at OAC, we used the Full Width at Half 
Maximum (FWHM) method. We applied this method to those orders of the echelle spectra which did not contain 
broad spectral lines affecting the shape of the cross-correlation function to compute the mean projected 
rotational velocities. As templates, we used a grid of artificially broadened spectra of non-rotating stars 
that have $T_{\rm eff}$, $\log g$, and $\rm [Fe/H]$ similar to the parameters of the program stars. 

For the hot stars observed at OAC and the stars observed with the CfA Digital Speedometers, we determined 
$v\sin i$ simultaneously with the atmospheric parameters when computing the best fit of the synthetic spectrum 
to the observed one. An upper limit of 5 and 2 km\,s$^{-1}$ in $v\sin i$ has been estimated according to the 
instrumental resolution of the spectrographs, as discussed in \citet{molenda07}. 

We find that one of the early-type stars, HIP\,94472, is rotating slowly $(v\sin i =13\,\rm km\,s^{-1})$
which makes it a promising target for asteroseismic modeling. The star might still be a fast rotator seen pole-on, 
but we remark that Bidelman classifies this star to the spectral type Am. We also found the abundance pattern
typical of Am stars that are slow rotators.  Thus, we conclude that HIP\,94472 is very likely a slowly rotating star.

The derived values of $v\sin i$ of our program stars are given in the last columns of Tables~\ref{teff}, 
\ref{model}, and \ref{model-dl}

\section{Discussion}
\subsection{The effective temperature}

\begin{figure}
\includegraphics[width=8.5cm,angle=0]{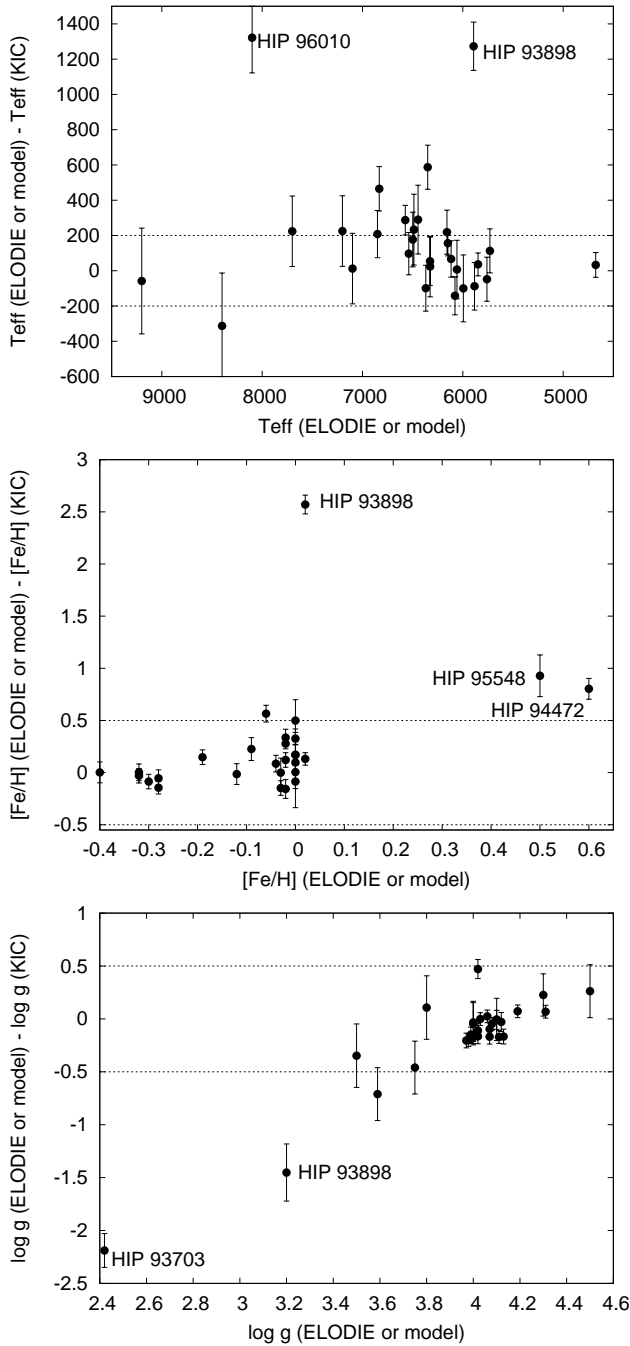}
\caption[]{The differences between 
$T_{\rm eff}$ ({\it top panel}), [Fe/H] ({\it middle panel}), and $\log g$
({\it bottom panel}) derived in this paper with the use of the reference stars from the ELODIE library (F, G,
and K stars) or from the model atmospheres (hot stars) and the values of $T_{\rm eff}$, [Fe/H] and $\log g$ 
in the Kepler Input Catalog. The dotted lines represent the stated uncertainty of the atmospheric parameters
in the KIC.}
\label{comparison}
\end{figure}

We compared the atmospheric parameters of the program stars derived in this paper with the determinations 
from the literature. For 11 stars form this paper, the effective temperature was derived from photometric indices by
\citet{nordstrom2004}, \citet{Masana2006}, \citet{Alonso1996}, \citet{Gonzalez2009}, \citet{Casagrande2006},
\citet{Fulbright2003}, \citet{Gray2003}, and \citet{Allende1999} (who derive $T_{\rm eff}$ from the comparison of 
the stars' position in the colour-magnitude diagram with the evolutionary models of solar metallicity); 
spectroscopic determinations by \citet{Robinson2007}, \citet{Soubiran2008}, and \citet{Tomkin1999} are available for 
seven stars. For most of the stars, $T_{\rm eff}$ in this paper and the values from the literature agree 
satisfactorily. Only for HIP\,97071, $T_{\rm eff}$ derived by \citet{nordstrom2004} and \citet{Masana2006} from
$uvby\beta$ and $JHK$ photometry is around 300\,K higher. Such a discrepancy, though, is not unusual taking into 
account that even a moderate error in the adopted $E(B-V)$ can imply an error of 100\,K or more in the photometric 
effective temperature \citep[see, e.g.,][]{Casagrande10}.

The comparison of $T_{\rm eff}$ in this paper with the values given in the Kepler Input Catalog (KIC) for 30 stars from 
our sample shows that in most cases the discrepancies do not exceed $\pm$200~K -- see Fig.~\ref{comparison}
where we plot the differences between the atmospheric parameters derived in this paper and the values listed in the KIC. 
The highest discrepancies occur for HIP\,93898 and HIP\,96010 for which the 
KIC lists $T_{\rm eff}$ lower by around 1300\,K. For the former star, also $\log g$ and [Fe/H] in this paper are 
significantly different from the values provided in the KIC. 

The reason for so large discrepancies in the atmospheric parameters of HIP\,93898 may be the fact that this is a close 
binary separated by $0^{\prime\prime}.5$. The difference of brightness between the components is 2.5 mag. Given the 
proximity of the stars, HIP\,93898 is very likely a physical binary. Thus, the secondary component may be significantly
cooler and redder than the primary. This could affect the broad-band photometry spanning a wide wavelength range, like
that used for deriving the atmospheric parameters in the KIC but the contamination of the optical spectrum should be 
marginal and its influence on the atmospheric parameters derived in this paper, negligible. This conclusion is 
supported by the fact that $T_{\rm eff}$ derived by us using the grid of \citet{moon85} and the narrow-band Str{\"o}mgren 
indices provided by \citet{hauck90}, 5797\,K, agrees well with the spectroscopic value.

We note also that HIP\,93898 falls in a very particular position in the HR diagram (see Fig.~\ref{hrdiagram}) at which the evolution 
is very fast. Since this region of the HR diagram is sparsely populated, stars discovered to fall in that location are very 
interesting objects which deserve further study.

The reason why our value of $T_{\rm eff}$ derived for HIP\,96010 is significantly higher than in the KIC is not clear, and the
literature does not provide photometric measurements which could be used to confirm the effective temperature derived 
in this paper. We note, though, that also this star is listed as an astrometric binary by \citet{Makarov05}.

\subsection{The metallicity}

For 16 stars from our sample, the metallicity was derived either from photometry by 
\citet{nordstrom2004}, \citet{Ibukiyama2002}, \citet{Alonso1996}, \citet{Zakhozhaj1996}, and \citet{Gray2003}
or from spectroscopy by \citet{Robinson2007}, \citet{Soubiran2008}, \citet{Fulbright2003},
\citet{Tomkin1999}, \citet{Thevenin1999}, and \citet{Peterson1981}.  All these derivations
agree with the values derived in this paper to within 1-2$\sigma$ error bars. The highest discrepancy
occurs for the subdwarf HIP\,94931 for which we derive the mean [Fe/H] $=-0.30$ dex which agrees with 
[Fe/H] $= -0.29$ derived from photometry by \citet{Ibukiyama2002} but not with the value derived by
\citet{Peterson1981} from high-resolution spectroscopy, [Fe/H]$= -0.74\pm0.13$.

The comparison of [Fe/H] in this paper with those given in the KIC shows that for most stars the differences
do not exceed $\pm 0.5$ dex, which is the stated uncertainty of the values in the KIC. The highest 
discrepancies occur for HIP\,93898 (2.6 dex) which was discussed in the previous section, HIP\,95548 (1 dex), 
and HIP\,94472 (0.8 dex), classified by \citet{Bidelman1985} to the spectral type Am, which indeed shows an enhanced
pattern of metals in its photosphere in our analysis. We do not find any indications of duplicity of that last star 
suggested by the spectral type F+A: assigned by \citet{skiff10}, either from the radial velocity or from the 
spectrum appearance.

\subsection{The surface gravity}

The surface gravity was derived spectroscopically for ten of our program stars by
\citet{Robinson2007}, \citet{Soubiran2008}, \citet{Tomkin1999}, and \citet{Thevenin1999},
or from photometry by \citet{Allende1999}, \citet{Gray2003}, and \citet{Fulbright2003}.
For all these stars, the $\log g$ values agree with our determinations to within 1-2$\sigma$ error bars.

The differences in $\log g$ derived in this paper and those listed in the KIC do not exceed 0.2 dex for 
most stars, and for all but one fall into the range of the stated precision of $\log g$ in the KIC to 
within their $1\sigma$ error bars. The highest discrepancies occur for the already discussed star HIP\,93898, 
and for HIP\,93703 which turns out to be a giant, unlike the KIC determination. 

\section{Summary}

We derived the atmospheric parameters, the radial velocity, and the projected rotational velocity of 
44 stars selected as \Kepler asteroseismic targets from the stars proposed by A.F.\ and J.M-\.Z.\ 
in the first run of the \Kepler proposals; to 33 of these stars we assigned the MK types.

We discovered three double-lined spectroscopic binary systems, HIP\,94924, HIP\,95115, and HIP\,97321, and we
classify two other stars, HIP\,94112 and HIP\,96062, as suspected single-lined spectroscopic binaries. For one of the 
SB2 stars discovered in this paper, HIP\,97321, we provide the orbital solution. The other SB2
systems require more observations to derive their orbital periods and compute the orbital elements.

The values of the projected rotational velocity which we provide can be used as a guideline for making 
selection of long-term asteroseismic targets for the \Kepler space mission. Our measurements can, e.g., help 
reject very fast rotators for which an asteroseismic modeling is difficult.

The precision of the atmospheric parameters given in this paper will allow to derive the mean density of the 
stars and to compute the mass and radius of the stars with a precision better than 3\% when these parameters 
are combined with the large separation of the frequencies detected in the \Kepler magnitudes. The derived radii 
may be then used to constrain accurate dimensions of transiting planets \citep{jcd2007} and to refine the 
scientific output of the \Kepler mission \citep{borucki2009}. Since the stellar mean densities can be also
derived from the analysis of the light curve of the transiting planets, a comparison of the densities derived 
in these two completely independent ways will give a direct test of the physics involved. Then, our measurements 
allowed us to complete the information on the atmospheric parameters which is missing in the KIC
for 14 stars from our sample, and to verify the values of $T_{\rm eff}$, [Fe/H], and $\log g$ listed in the KIC 
for the remaining 30 stars. 

We found that although for most of the discussed stars $T_{\rm eff}$, $\log g$ and [Fe/H] in this paper and in 
the KIC agree to within the KIC uncertainties, for some the differences can be high. The stars for which we report 
the highest discrepancies between $T_{\rm eff}$, [Fe/H], and $\log g$ in this paper and in the KIC are 
HIP\,93703, HIP\,93898, HIP\,94472, HIP\,95548, and HIP\,96010.

Our results show that ground-based studies aiming at deriving atmospheric parameters of \Kepler asteroseismic
targets are crucial for the successful asteroseismic modeling of these stars, and for making a full use of the 
potential of the \Kepler space mission.

\section*{Acknowledgments}
This work has been partly supported by \Kepler mission under cooperation agreement NCC2-1390 (D.W.L., PI),
the Italian {\em Ministero dell'Istruzione, Universit\`a e Ricerca} (MIUR), and the Polish MNiSW grant
N203~014~31/2650 which are gratefully acknowledged.

\label{lastpage}

\end{document}